\documentclass[aps,pre,showpacs,showkeys,amssymb,twocolumn]{revtex4}

\usepackage{bm}
\usepackage{graphicx}
\usepackage{amsfonts}
\usepackage{amsmath}

\begin{document}

\title{Phase transitions triggered by dumbbell equipotential hypersurfaces}

\author{Fabrizio Baroni}

\email{baronifab@libero.it}
\affiliation{Formerly at Dipartimento di Fisica dell'Universit\`a di Firenze, Via G. Sansone 1, I-50019 Sesto F.no (FI), Italy}

\date{\today}

\begin{abstract}
In a recent paper a toy model (called hypercubic model) undergoing a first-order $\mathbb{Z}_2$ symmetry breaking phase transition (SBPT) has been introduced. The hypercubic model was inspired by the \emph{topological hypothesis}, according to which a phase transition may be entailed by suitable topological changes of the equipotential hypersurfaces $\Sigma_v$ of configuration space. The $\Sigma_v$'s of the hypercubic model have a single topological change, which, under further particular hypotheses of geometric nature, entails the $\mathbb{Z}_2$-SBPT. In this paper we introduce an extended version of the hypercubic model in which no topological change in the $\Sigma_v$'s is present anymore, but nevertheless the $\mathbb{Z}_2$-SBPT occurs the same. We introduce a geometric property of the $\Sigma_v$'s (i.e. dumbbell $\Sigma_v$'s suitably defined) that is sufficient to entail a $\mathbb{Z}_2$-SBPT regardless their topology. The paper ends by applying the picture of the dumbbell $\Sigma_v$'s to a physical model, i.e. the mean-field $\phi^4$ model.
\end{abstract}



\maketitle


\section{Introduction}

Phase transition are very common in nature. They are sudden changes of the macroscopic behavior of a natural system composed by many interacting parts occurring while an external parameter is smoothly varied. Phase transitions are an example of emergent behavior, i.e. of a collective properties having no direct counterpart in the dynamics or structure of individual atoms \cite{lebowitz}. The successful description of phase transitions starting from the properties of the interactions between the components of the system is one of the major achievements of equilibrium statistical mechanics.

From a statistical-mechanical point of view, in the canonical ensemble, describing a system at constant temperature $T$, a phase transition occurs at special values if the temperature called transition points, where thermodynamic quantities such as pressure, magnetization, or heat capacity, are non-analytic functions of $T$; these points are the boundaries between different phases of the system. Starting from the celebrated solution of the $2D$ Ising model by Onsager, these singularities have been indeed found in many models, and later developments like the renormalization group theory \cite{goldenfeld} have considerably deepened our knowledge of the properties of the transition points, at least in the case continuous transitions, or critical phenomena.

Yet, the situation is not completely satisfactory, First, in the canonical ensemble these singularities occur only in the rather artificial case of infinite systems: following an early suggestion by Kramers \cite{c}, Lee and Yang \cite{yl} showed that in the thermodynamic limit $N\rightarrow\infty$ ($N$ is the number of degrees of freedom, and the limit is taken at fixed density) must be invoked to explain the existence of true singularities in the canonical partition function $Z(T)$ and then in the thermodynamic functions defined as derivatives of $Z(T)$. Since in the last decades many examples of transitional phenomena in systems far form the thermodynamic limit have been found (e.g. in nuclei, atomic clusters, biopolymers, superconductivity, superfluidity), a description of phase transition valid also for finite systems would be desirable. Second, while necessary conditions fro the presence of a phase transitions can be found (one example is the above-mentioned need of the thermodynamic limit in the canonical ensemble), nothing general is known about sufficient conditions: no general procedure is at hand to tell if a system where a phase transition is not ruled out from the beginning does have or not such a transition without computing $Z$. This might indicate that our deep understanding of this phenomenon is still incomplete. 

These considerations motivate a study of the deep nature of phase transitions which may also be based on alternative approaches. One of such approaches, proposed in \cite{cccp} and developed later \cite{ccp}, is based on simple concepts and tools drawn from differential geometry and topology. The main issue of this new approach is a topological hypothesis, whose content is that at their deepest level phase transitions are due to a topology change of suitable submanifolds of configuration space, those where the system lives as the number of its degrees of freedom becomes very large.

This idea has been discussed and tested in many recent papers \cite{acprz,b,bc,ccp1,ckn,gss,gm,k,rs}. Moreover, the topological hypothesis has been given a rigorously background by a theorem \cite{fp} which states that, at least for systems with short-ranged interactions and confining potentials, topology changes in configurations space submanifolds are a necessary condition for a phase transition. Anyway, in \cite{km} it has been shown that the theorem cannot be sustainable except for some particular cases. In \cite{gfp} a generalized version of the theorem has been proposed in which also concepts of geometric nature are encompassed besides topology. However, the problem of which may be, if any, the sufficient conditions to entail phase transitions either topological alone or in addition to geometric hypotheses remains open. A first answer to this problem has been given in \cite{bc}, where a straightforward theorem for the occurrence of a $\mathbb{Z}_2$-symmetry breaking phase transition (SBPT hereafter) has been proven. 

In this paper we generalize that theorem finding out a more general sufficient and necessary condition for $\mathbb{Z}_2$-SBPTs. The sufficient condition of the theorem in \cite{bc} makes use only of topological properties of the equipotential hypersurfaces ($\Sigma_v$'s hereafter) of configuration space, while the generalized version is given in terms of a geometric property of the $\Sigma_v$'s largely independent on their topology. The $\Sigma_v$'s having this property has been called \emph{dumbbell-shaped} in the sense of the presence of a neck which will be clarified in the paper. The original topological condition survives as a limiting case. In general, the results of this paper seems to suggest that purely topological conditions for SBPTs are unattainable, as already highlighted in \cite{k1}, even though it is not said that the generating-mechanism of SBPTs delineated here is the unique acting in nature. 

The paper is structured in a pedagogical way. In Sec. \ref{hcm} we start by introducing an extended version of a toy model (called \emph{hypercubic model}) undergoing a first-order $\mathbb{Z}_2$-SBPT introduced in \cite{bc}, showing in an intuitive way how the mechanism of the dumbbell $\Sigma_v$'s works. In Sec.  \ref{strangled} we give a rigorous definition of what a dumbbell $\Sigma_v$ is, and we generalize the theorem in \cite{bc}. In Sec. \ref{bs} we introduce a model based on the hypercubic model which rigorously shows what depicted in Sec. \ref{hcm} from a mathematical point of view. Finally, in Sec. \ref{phi4} we apply the framework of the dumbbell $\Sigma_v$'s to a physical model, i.e. the mean-field $\phi^4$ model.

\section{Hypercubic model}
\label{hcm}

\begin{figure}
\begin{center}
\includegraphics[width=0.235\textwidth]{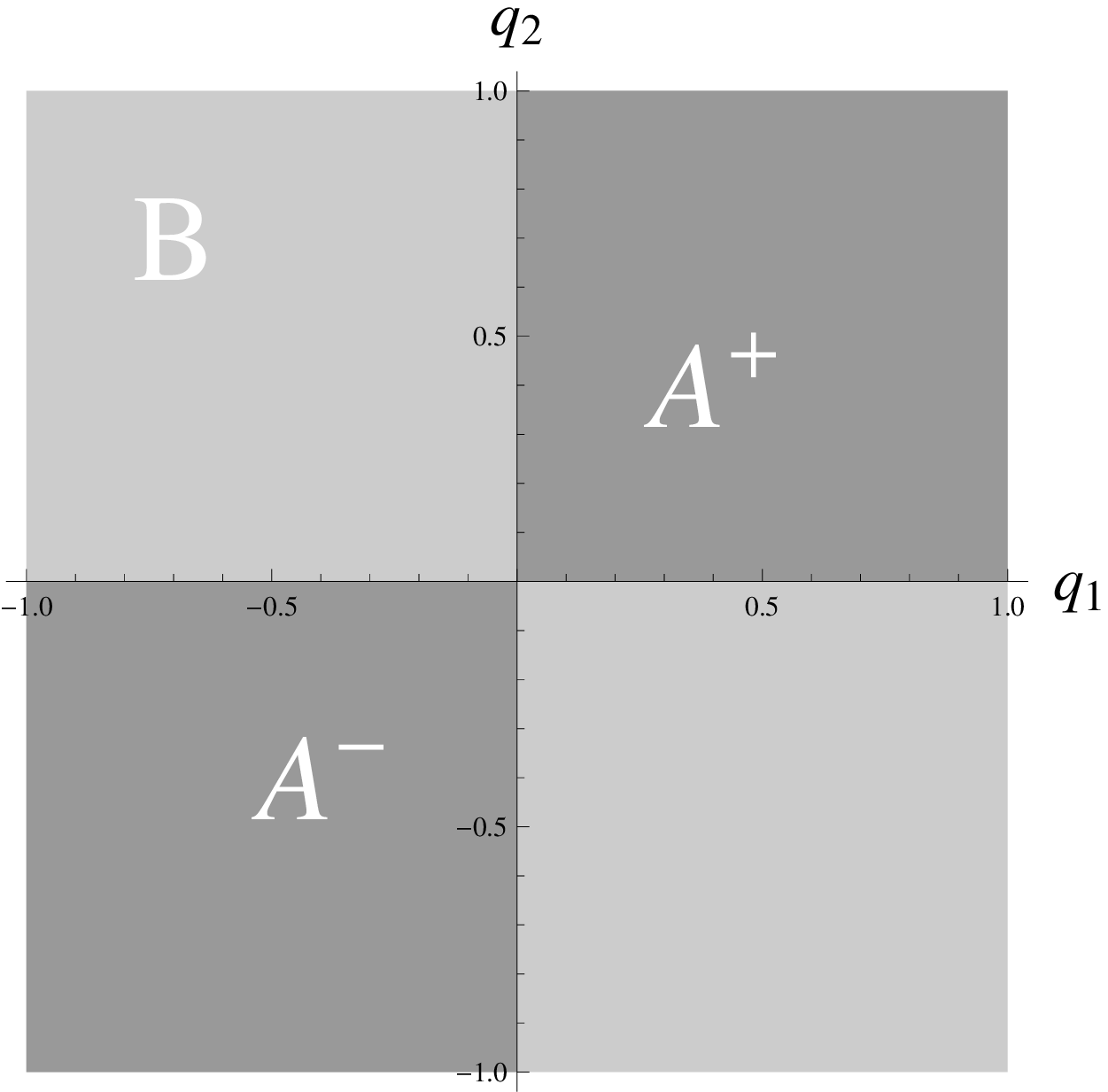}
\includegraphics[width=0.235\textwidth]{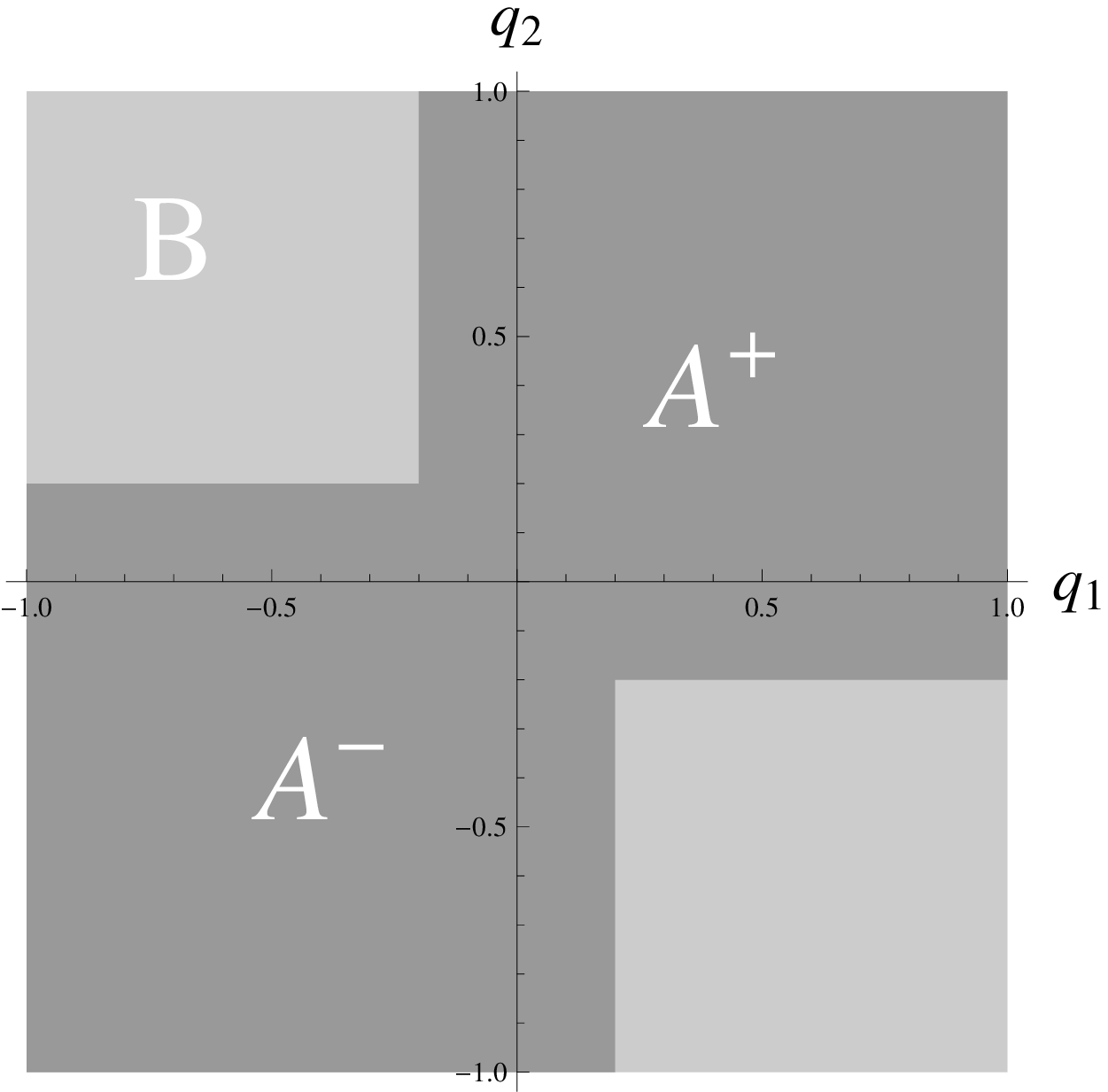}
\caption{Sketch of the $\Sigma_v$'s (\ref{}) of the hypercubic model (\ref{Vhcm}) for $N=2$. The sides of the $2$-cubes (squares) $B$, $A^+$, $A^-$ are $b=2$, $a=1$, $a=1$ respectively, as defined in \cite{bc} (left), while $b=2$, $a=1.2$, $a=1.2$ respectively, as modified in Sec. \ref{hcm}.}
\label{cubes}
\end{center}
\end{figure}

In this section we will briefly recall the toy model, called hypercubic model, introduced in \cite{bc}. The model has been defined in such a way to undergo a first-order $\mathbb{Z}_2$-SBPT. The potential is nothing but a generalization to $N$ dimensions of a 
$\mathbb{Z}_2$-symmetric square double-well potential with the gap between the wells proportional to the number of degrees of freedom $N$, that is
\begin{equation}
V(\textbf{q})=\left\{\begin{array}{ll}
-N v_c\quad &\hbox{if} \quad\textbf{q}\in A^{\pm}
\\
0 &\hbox{if} \quad\textbf{q}\in B\backslash\{A^+\cup A^- \}
\\
+\infty &\hbox{if} \quad\textbf{q}\in \mathbb{R}^N\backslash B
\end{array}\right.,
\label{Vhcm}
\end{equation}
where $v_c>0$, $B$ is an $N$-cube of side $b$ centered in the origin of coordinates, and $A^+, A^-$ are $N$-cubes of side $a$ disposed in such a way to be included in $B$ and to be one the image of the other under the $\mathbb{Z}_2$ symmetry. Furthermore, $A^+, A^-$ are disposed in such a way to be at the maximum from each other (see Fig. \ref{cubes}), and $b\geq 2a$ has been assumed to have $A^+\cap A^-=\emptyset$.

The equipotential hypersurfaces of configuration space are defined as follows
\begin{equation}
\Sigma_{v,N}=\{\mathbf{q}\in\mathbb{R}^N: v(\mathbf{q})=v\},
\label{sigmavN}
\end{equation}
where $v=\frac{V}{N}$. The $\Sigma_{v,N}$'s of the hypercubic model are the following
\begin{equation}
\Sigma_{v,N}=\begin{cases}
 \emptyset& \text{ if } v<-v_c 
\\ 
 A^+\cup A^-& \text{ if } v=-v_c 
\\ 
 \emptyset& \text{ if } -v_c<v<0
\\ 
 B& \text{ if } v=0 
\\ 
 \emptyset& \text{ if } v>0
\end{cases},
\end{equation}
thus, a topological change occurs as the potential jumps between $-v_c$ and $0$. 


The canonical partition function is
\begin{eqnarray}
Z_N(T)&=&\int_{\mathbb{R}^N}dq\,e^{-\frac{V(\mathbb{q})}{T}}=\nonumber
\\
&=&\int_{A^+\cup A^-}dq\,e^{\frac{Nv_c}{T}}+\int_{B\backslash\{A^+\cup A^- \}}dq\,=\nonumber
\\
&=&2 a^N\,e^{\frac{N v_c}{T}}+(b^N-2a^N).
\end{eqnarray}
As $N$ is large enough, $Z_N$ can be approximated as follows

\begin{equation}
Z_N\simeq 2e^{N\left(\ln a+\frac{v_c}{T}\right)}+e^{N\ln b}.
\end{equation}
In the limit $N\rightarrow\infty$ the critical temperature $T_c=\frac{v_c}{\ln(b/a)}$ arises, because as $T<T_c$ the first addendum in the right hand side of the last equation survives, while as $T>T_c$ the second addendum survives.

In the thermodynamic limit the free energy, the average potential, and the specific heat are respectively

\begin{equation}
f=-\frac{T}{N}\ln Z_N=\left\{\begin{array}{ll}
-T\ln a-v_c   & \hbox{if} \quad T\le T_c
\\
-T\ln b   & \hbox{if} \quad T\ge T_c
\end{array}\right.,
\end{equation}

\begin{equation}
\left\langle v\right\rangle=-T^2\frac{\partial}{\partial T}\left(\frac{f}{T}\right)=\left\{\begin{array}{ll}
-v_c & \hbox{if} \quad T<T_c
\\
0 & \hbox{if} \quad T>T_c
\end{array}\right.,
\end{equation}

\begin{equation}
C_v=\frac{\partial \left\langle v\right\rangle}{\partial T}=\left\{\begin{array}{ll}
0 & \hbox{if} \quad T<T_c
\\
+\infty  & \hbox{if} \quad T=T_c
\\
0 & \hbox{if} \quad T>T_c.
\end{array}\right..
\end{equation}
The spontaneous magnetization in the broken phase has been assumed to be the center of mass coordinates of the $N$-cubes $A^+$ or $A^-$
\begin{equation}
\left\langle m\right\rangle=\left\{\begin{array}{ll}
\pm\frac{b-a}{2} & \hbox{if} \quad T<T_c
\\
0 & \hbox{if} \quad T>T_c.
\end{array}\right.,
\end{equation}
Summarizing, the model shows the complete picture of a first-order $\mathbb{Z}_2$-SBPT.

\begin{figure}
\begin{center}
\includegraphics[width=0.235\textwidth]{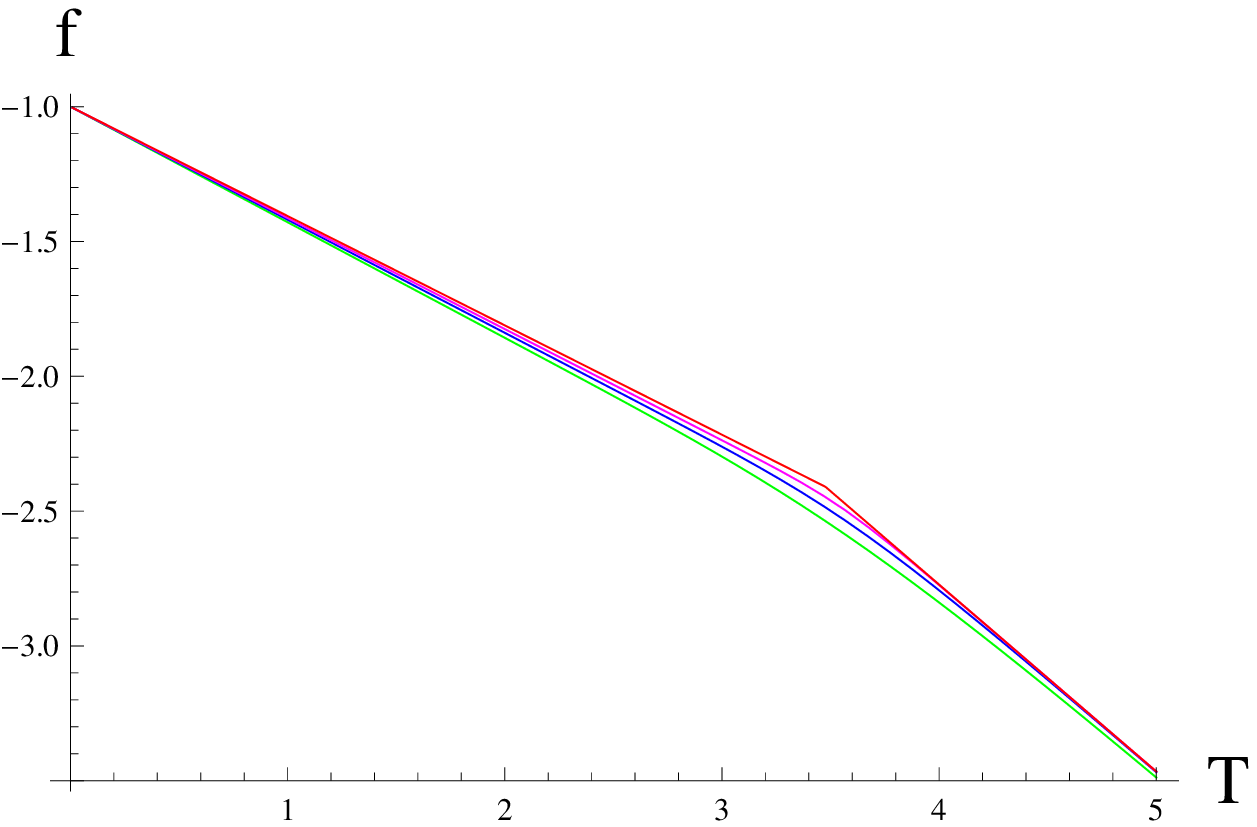}
\includegraphics[width=0.235\textwidth]{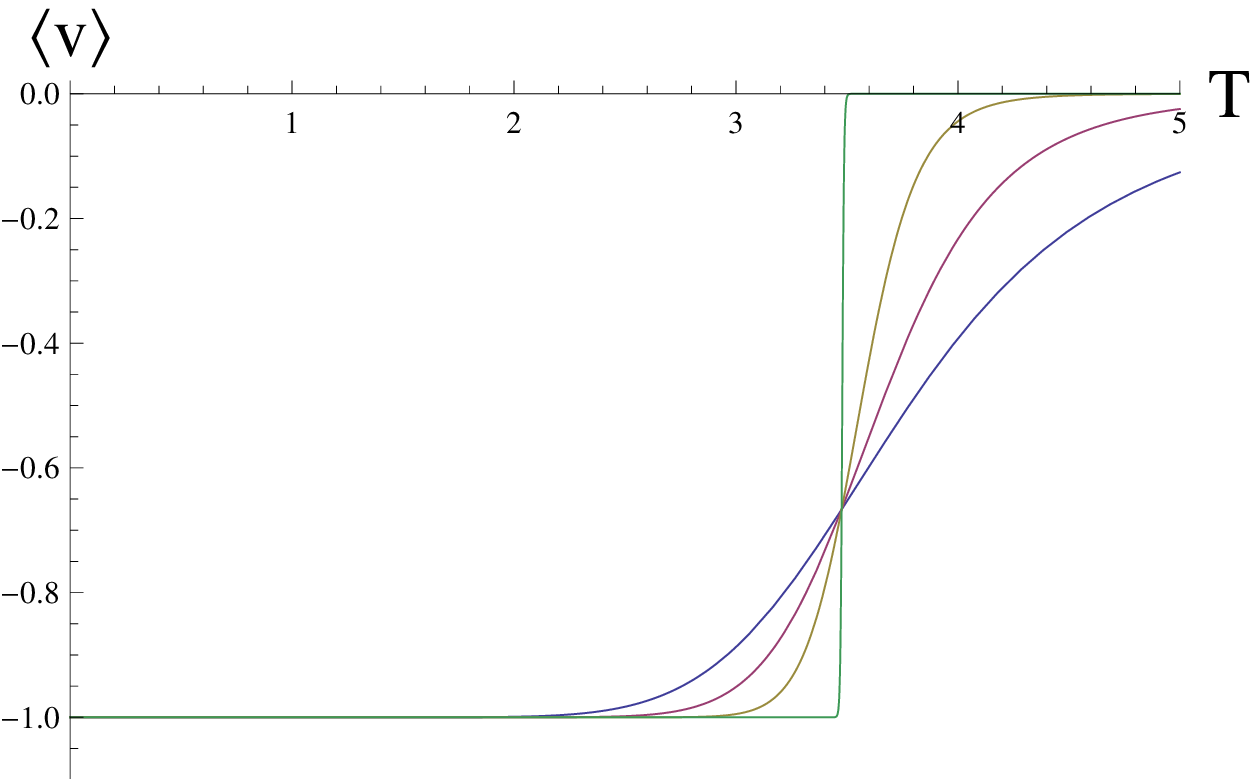}
\includegraphics[width=0.235\textwidth]{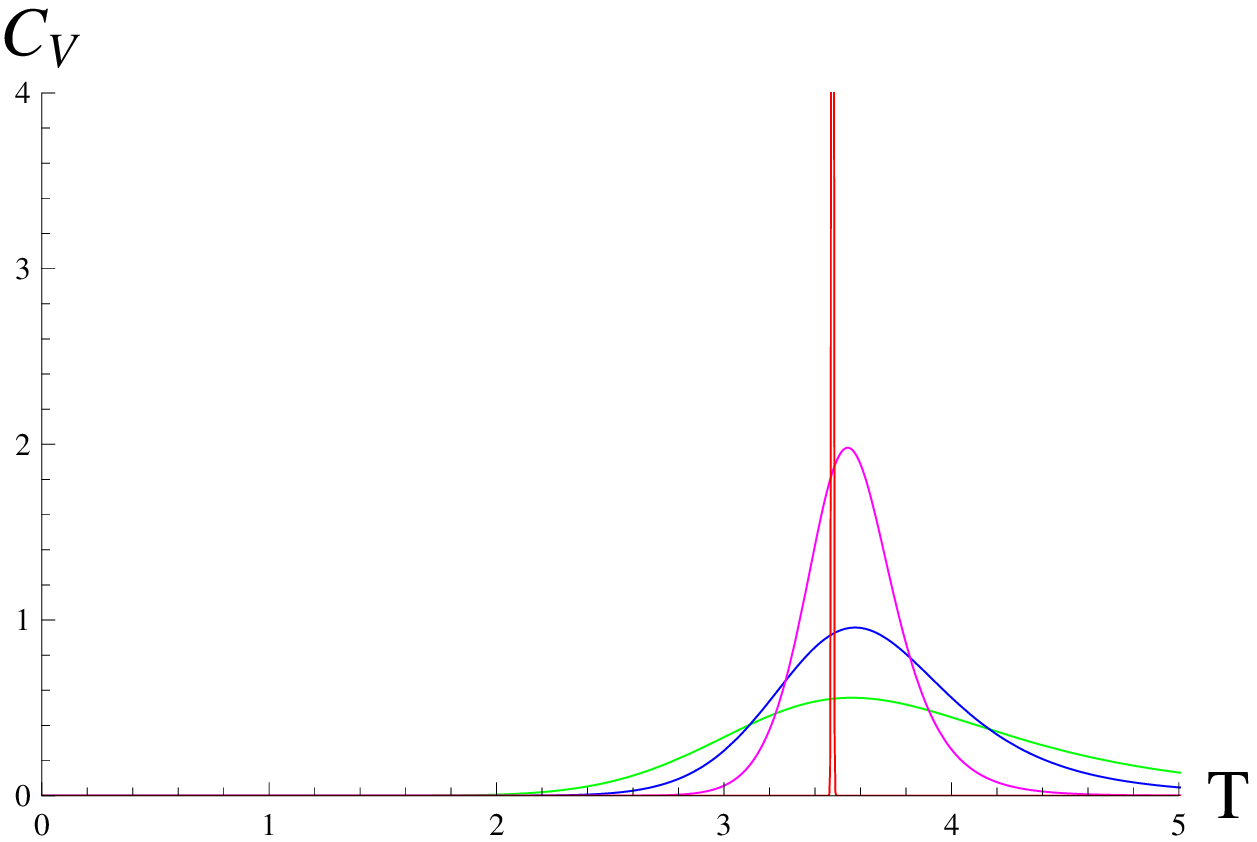}
\includegraphics[width=0.235\textwidth]{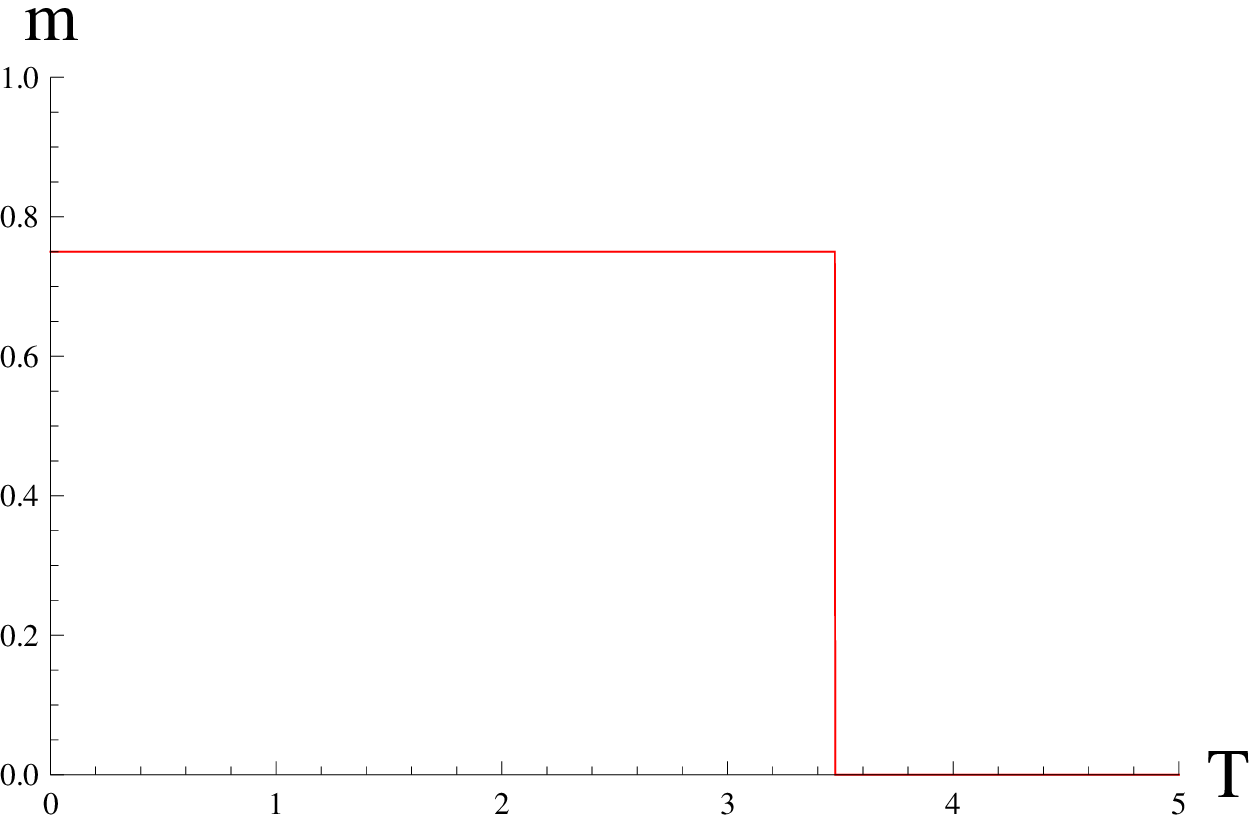}
\caption{Hypercubic model (\ref{Vhcm}) with $v_c=1$ $a=1.5$, and $b=2$. From top to bottom, and from left to right: free energy $f$, average potential $\left\langle v\right\rangle$, specific heat $C_v$, and magnetization $\left\langle m\right\rangle$. The smooth lines are for $N=30, 50, 100$ (green, blue, magenta), and show a non-uniform convergence toward a discontinuous limit for $N\to\infty$ (red) corresponding to a first-order phase transition with critical temperature $T_c=\frac{1}{\ln(4/3)}$.}
\label{}
\end{center}
\end{figure}

In \cite{bc} the authors have assumed $b\ge 2a$ on the $N$-cubes sides, because they were guided by the idea that in order to entail a SBPT at least a topological change in the $\Sigma_{v,N}$'s is needed. The idea is that, if the $N$-cubes $A^+$, $A^-$ are disjoint, the probability of the representative point (RP hereafter) to jump between $A^+$, and $A^-$ is vanishing in the thermodynamic limit. But here we will show that the disjointness of $A^+$, $A^-$ is not necessary, and that a SBPT can be entailed by another condition on $A^+\cup A^-$ which includes the disjointness as a particular case.

We start by observing that the solution of the thermodynamic does not require the restriction $b\ge 2a$, and it makes sense also under the less restrictive condition $b>a$. Indeed, $T_c=\frac{v_c}{\ln(b/a)}>0$ needs only $b>a$, and if $b=a$, $T_c=0$, and the SBPT disappears. But as $b>a>b/2$, $A^+\cap A^-\neq\emptyset$. Thus, how can the $\mathbb{Z}_2$ symmetry be broken as $T<T_c$? And, what happens to the spontaneous magnetization?  Assuming the 'a priori' equal probability hypothesis, the spontaneous magnetization should be vanishing because $A^+\cup A^-$ is symmetric under $\mathbb{Z}_2$. 

To help the intuition, consider the case in which $a$ is only slightly greater than $b/2$. $A^+\cap A^-$ is a small $N$-cube of side $2a-b$. How can the RP go freely around the whole $A^+\cup A^-$ if it has to pass through $A^+\cap A^-$? It is clear that the $\mathbb{Z}_2$ symmetry has to break even if $A^+\cap A^-\neq\emptyset$. In what follows we will see that it is not important how small $A^+\cap A^-$ is, but that the crucial feature of $A^+\cup A^-$ is that it is 'strangled' in the sense that will be cleared. 

More precisely, we are in front of the spontaneous ergodicity breaking phenomenon, that among its consequences, includes symmetry breaking. We recall that the ergodic hypothesis implies that we can substitute the time average of an observable $F(\mathbf{q})$ of the system, i.e. a function of the coordinates, by the canonical (or microcanonical) ensemble average
\begin{equation}
\overline{F}=\lim_{t\rightarrow+\infty}\frac{\int^{t}_{0}dt \,F(\mathbf{q}(t))}{t}=\left\langle F\right\rangle=\frac{\int_M d\mathbf{q}\, F(\mathbf{q}) \,e^{-\beta V(\mathbf{q})}}{Z(\beta)},
\label{}
\end{equation}
where $M$ is the configuration space.

\begin{figure}
\begin{center}
\includegraphics[width=0.235\textwidth]{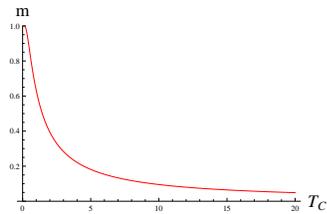}
\caption{Hypercubic model (\ref{Vhcm}) with $v_c=1$, and $b=2$. Spontaneous magnetization $m=(b-a)/2$ as a function of the critical temperature $T_c=\frac{v_c}{\ln(b/a)}$. Only the positive branch is plotted.}
\end{center}
\end{figure}

\section{Dumbbell equipotential hypersurface of configuration space}
\label{strangled}

Consider an $N$ degrees of freedom system. Let us define the hyperplane of $\mathbb{R}^N$ at constant magnetization
\begin{equation}
\pi_{m,N}=\{\mathbf{q}\in\mathbb{R}^N: \frac{1}{N}\sum^{N}_{i=1}q_i=m\},
\label{pim}
\end{equation}
and define the function
\begin{equation}
a_N(v,m)=\omega_N(v,m)^\frac{1}{N},
\end{equation}
where 
\begin{equation}
\omega_N(v,m)=\mu(\Sigma_{m,N}\cap\pi_{m,N})=\int_{\Sigma_{m,N}\cap\pi_{m,N}}\frac{d\Sigma}{\|\nabla V\|}
\end{equation}
is the microcanonical volume, or density of states, at fixed $v$ and $m$. $a_N(v,m)$ is linked to the microcanonical entropy $s_N(v,m)$ by the relation 
\begin{equation}
s_N(v,m)=\ln a_N(v,m).
\end{equation}

\smallskip
\noindent{\textbf{Definition.}} \emph{A $\Sigma_{v,N}$ is dumbbell-shaped if $a_N(v,m)$, or equivalently $s_N(v,m)$, does not take the maximum at $m=0$.}

\smallskip
Knowing where the maximum of $a_N(v,m)$ is located is crucial to determine the spontaneous magnetization $\left\langle m\right\rangle$. Indeed, the last takes the value of $m$ at which $a_N(v,m)$ is maximum. Let us see why. Consider the co-area formula \cite{pettini} of the canonical partition function
\begin{equation}
Z_N=N\int_{v_{mim}}^{+\infty}dv \,e^{-N\beta v}\mu\left(\Sigma_{v,N}\right),
\end{equation}
where $v_{mim}$ is the absolute minimum of the potential density. Since $\Sigma_{v,N}=\cup_{m\in\mathbb{R}}\left(\Sigma_{v,N}\cap\pi_{m,N}\right)$
\begin{eqnarray}
\mu\left(\Sigma_{v,N}\right)=\sqrt{N}\int dm\,\mu(\Sigma_{m,N}\cap\pi_{m,N})\nonumber
\\
=\sqrt{N}\int dm\,e^{N s_N(v,m)}.
\end{eqnarray}
In the thermodynamic limit \footnote{For the sake of precision, even though we consider the thermodynamic limit, $N$ must be considered finite although very large, because the definition of $\Sigma_{v,N}$ (\ref{sigmavN}) does not make sense anymore if $N\rightarrow\infty$. The situation is similar to what happens in the infinitesimal calculus, where we consider infinitesimal quantities, like $dx$, which are small how much we want, but are nonzero.} $Z_N$ can by evaluated by the saddle point approximation, so that
\begin{equation}
\mu\left(\Sigma_{v,N}\right)\propto e^{N s_N(\left\langle v\right\rangle,m_0)},
\end{equation}
where $\left\langle v\right\rangle$ is the average potential, and $m_0$ maximizes $s_N(\left\langle v\right\rangle,m)$. 

Now it is clear that the $\Sigma_{\left\langle v\right\rangle,N}$ to be a dumbbell one is a sufficient condition for $\mathbb{Z}_2$-symmetry breaking (SB). Furthermore, this condition appears also as necessary, because if the $\Sigma_{\left\langle v\right\rangle,N}$ were not dumbbell-shaped, then the maximum of $a_N(\left\langle v\right\rangle,m)$ would be at $m=0$ so that the $\mathbb{Z}_2$ symmetry would not be broken, against the hypothesis. These consideration, for the sake of mathematical formality, can be condensed in a straightforward theorem.

\smallskip
\noindent{\textbf{Theorem.}} \emph{Let $(v',v'')$ be an interval of accessible values of the potential density of an $N$ degrees of freedom Hamiltonian system with a $\mathbb{Z}_2$ symmetry. The $\mathbb{Z}_2$ symmetry is spontaneously broken for suitable values of the temperature $T$ such that $\left\langle v\right\rangle(T)\in (v',v'')$ if, and only if, there exists $N_0\in \mathbb{N}$ such that the $\Sigma_{v,N}$'s are dumbbell-shaped $\forall N>N_0$ and $\forall v\in(v',v'')$.}

\smallskip
In the most common case $v'$ is the absolute minimum of the potential density, which is reached at $T=0$. If a phase transition, meant as the transition point between the broken phase and the unbroken one, is associated to the SB, it is a value of the potential $v_c$ that separates the dumbbell $\Sigma_{\left\langle v\right\rangle,N}$'s from the ones which are not dumbbell-shaped. This means that there exists an $N_0$ such that $\forall N>N_0$ the $\Sigma_{v,N}$'s are dumbbell-shaped for $v>v_c$, and are not dumbbell-shaped for $v<v_c$. $\Sigma_{v_c,N}$ plays the role of a sort of critical $v$-level set. 

At this point it is worth distinguishing two cases in the broken phase and in the thermodynamic limit.

\smallskip
(i) $a(v,m)=\lim_{N\rightarrow\infty}a_N(v,m)$ is a non-concave function of $m$ with two absolute maxima corresponding to values of the spontaneous magnetization. Since $a(v,m)$ has to be concave for short-range potentials \cite{g,l}, this is the picture of a long-range potential.

\smallskip
(ii) $a(v,m)$ does not maintain the non-concavity at finite $N$ of the $a_N(v,m)$'s, as a consequence is a non-strictly concave function of $m$. This is the typical picture of a short-range system, even though a long-range one cannot be 'a priori' excluded.

\smallskip
To make clear the situation we give a physical example. Consider a ferromagnetic material, which can be modeled by a short-range $3D$ Ising model. As the temperature is above the Curie temperature $T_c$ the $O(3)$ symmetry is unbroken, while as the temperature is below $T_c$ the $O(3)$ symmetry is broken. Anyway, the modulus of the spontaneous magnetization $\left\langle m\right\rangle$ can take every value between zero and its maximum, because the free energy, or equivalently the entropy, is a flat function of $m$ in the same interval. This is the picture of case (ii). 

\begin{figure}
\begin{center}
\includegraphics[width=0.235\textwidth]{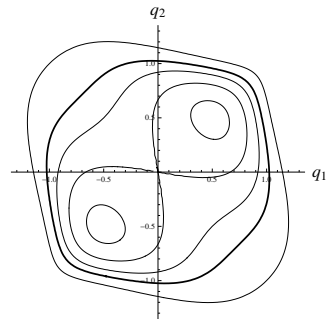}
\caption{Some $\Sigma_{v,2}$'s of a $\mathbb{Z}_2$-symmetric double-well potential for increasing values of $v$ starting from the innermost one. The thick $\Sigma_{v,2}$ is critical in the sense that separates the dumbbell ones from that which are not dumbbell-shaped. In two dimensions this can be only intuited, because the intersection with the $\pi_{m,2}$'s is made by only two points, but in more dimensions it is a $(N-2)$-submanifold of $\mathbb{R}^N$.}
\end{center}
\label{dwp}
\end{figure}

We wonder what may be a general sufficient condition for $\mathbb{Z}_2$-SBPTs given on the potential globally considered. As it has already pointed out in \cite{b1}, a double-well potential may be the most general answer, e.g. the $\phi^4$ model in Sec. \ref{phi4}. The wells of the potential have to be located on a line orthogonal to the $\pi_{m,N}$'s defined in (\ref{pim}). This scenario is represented in Fig. \ref{dwp}. Following the example of the hypercubic model, an alternative way is to define the potential foil by foil by shaping the single $\Sigma_{v,N}$ as we want, but this way we cannot guarantee the smoothness of the potential or any other analytical property. 

\subsection{On a theorem on a sufficient condition for $\mathbb{Z}_2$-SBPT}

In \cite{bc} a straightforward theorem (theorem 1) on a sufficient topological condition for $\mathbb{Z}_2$-SBPT has been proven. It is given starting from the topological properties of the $\Sigma_{v,N}$'s of an $N$ degrees of freedom Hamiltonian system with a $\mathbb{Z}_2$ symmetry. Simplifying a little bit the scenario, the statement is as follows.

Let $v'<v''$ be two values of the potential density such that $\Sigma_{v,N}=\Sigma_{v,N}^A\cup\Sigma_{v,N}^B$ $\forall v\in (v',v'')$, $\Sigma_{v,N}^A\cap\Sigma_{v,N}^B=\emptyset$, $\Sigma_{v,N}^A\sim\Sigma_{v,N}^B\sim\mathbb{S}^N$ where '$\sim$' stands for 'is homeomorphic to', and such that $\Sigma_{v,N}^A$ is the image of $\Sigma_{v,N}^B$ under the $\mathbb{Z}_2$ symmetry. Then, in the thermodynamic limit the $\mathbb{Z}_2$ symmetry is spontaneous broken for the values of temperature $T\in (T',T'')$, where $v'=\left\langle v\right\rangle(T')$ and $v''=\left\langle v\right\rangle(T'')$.

In that theorem it has been made the assumption that if the RP is confined in one of the two connected components of $\Sigma_{v,N}$ the spontaneous magnetization can be calculated by the ensemble average on the connected component itself. At the light of what discussed here this is trivially not always true. Nevertheless, the theorem in \cite{bc} survives as a particular case of the theorem given in Sec. \ref{strangled}, if we add the hypothesis that $\Sigma_{v,N}$ is symmetric with respect to the plane $\pi_0$ defined in (\ref{pim}). Indeed, since $\Sigma_{v,N}$ is made by two disjoint connected components $\mathbb{Z}_2$-non-symmetric singularly considered, then $\Sigma_{v,N}\cap\pi_0=\emptyset$, so that $\Sigma_{v,N}$ is 'strangled' and the $\mathbb{Z}_2$ symmetry is broken because of theorem in Sec. \ref{strangled}.

\subsection{On a theorem on a necessary condition for phase transitions}
\label{pettini}

\begin{figure}
\begin{center}
\includegraphics[width=0.235\textwidth]{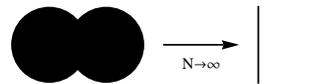}
\caption{The sketch illustrates a dumbbell $\Sigma_{v,N}$ whose 'topological limit' for $N\rightarrow\infty$ is different from the topology at any $N$ in the sense explained in Sec. \ref{pettini}.}
\label{limtop}
\end{center}
\end{figure}

In \cite{fps,fp,pettini} a theorem on a necessary condition for phase transitions, meant as a loss of analyticity in the free energy, has been proven. The necessary condition is a topological change in the $\Sigma_{v,N}$'s located in correspondence to the average potential $v_c=\left\langle v\right\rangle(T_c)$, where $T_c$ is the critical temperature. Despite that in \cite{km} a counterexample has been found, i.e. the $2D$ $\phi^4$ model, the original idea of the theorem, and more in general of the \emph{topological hypothesis} in \cite{pettini}, may be in somewhat extent recovered by considering a sort of topological limit for $N\rightarrow\infty$ of the $\Sigma_{v,N}$'s, as sketched in Fig. \ref{limtop}.

Here, it has clearly showed how a SBPT arises as an effect of the restriction of the integration domain in configuration space of the canonical partition function in the limit $N\rightarrow\infty$. As $N$ becomes larger an larger, the canonical measure shrinks more and more around $\Sigma_{v_c,N}$, so that the integration domain does the same, but even though the topology of $\Sigma_{v_c,N}$ can be trivially that of an $N$-sphere, $\Sigma_{v_c,N}$ may be 'strangled' giving rise to the SBPT as a consequence. We may say that the 'topological limit' for $N\rightarrow\infty$ of the $\Sigma_{v,N}$'s is different from the topology of the $\Sigma_{v,N}$ themselves. Anyway, this is only an imaginative way of thinking, because in the limit $N\rightarrow\infty$ no $\Sigma_{v,N}$ exists anymore.

In \cite{gfp} the authors of the theorem has proposed a generalized version of it, which includes the above-mentioned counterexample, based on the new concept of asymptotic diffeomorphicity, according to which two manifolds can be diffeomorphic at any $N$ but not anymore in the limit $N\rightarrow\infty$. While waiting for a rigorous definition of asymptotic diffeomorphicity, we wonder what it may have in common, if any, with the concept of the 'strangled' manifolds introduced here for the $\mathbb{Z}_2$ symmetry.

\smallskip
\noindent{\emph{Remark.}} The fact that a $\Sigma_{v,N}$ is 'strangled' is not related to its topology at all. A 'strangled' $\Sigma_{v,N}$ can have any topology.

\section{Toy model derived from hypercubic model}
\label{bs}

\begin{figure}
\begin{center}
\includegraphics[width=0.235\textwidth]{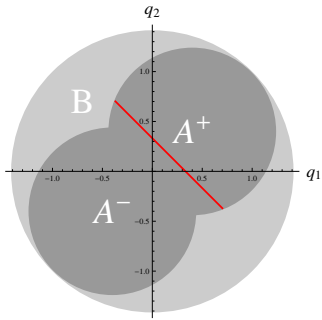}
\caption{Sketch for $N=2$ of the $\Sigma_{v,N}$'s of the model in Sec. \ref{bs}: $\Sigma_{-v_c,2}=A^+\cup A^-$ and $\Sigma_{0,2}=B\backslash\left(A^+\cup A^-\right)$ where $B$ is the whole $2$-ball. The red segment is $A^+\cup A^-\cap\pi_{m,2}$.}
\label{ballesecanti}
\end{center}
\end{figure}

Unfortunately, the calculation of $A^+\cup A^-\cap\Sigma_{m,N}$ where $\Sigma_{m,N}$ of the hypercubic model (\ref{Vhcm}) is not so easy. 
Therefore, we replace the $N$-cubes by $N$-balls, as sketched in Fig. \ref{ballesecanti}. The radius of $B$ is assumed $\sqrt{N}$ to give rise to a magnetization $m\in [-1,1]$, while the radius of $A^+$, $A^-$ is $(1-m_0)\sqrt{N}$, where $m_0$ is the value of the spontaneous magnetization. $0<m_0<1$ is assumed.

\begin{figure}
\begin{center}
\includegraphics[width=0.235\textwidth]{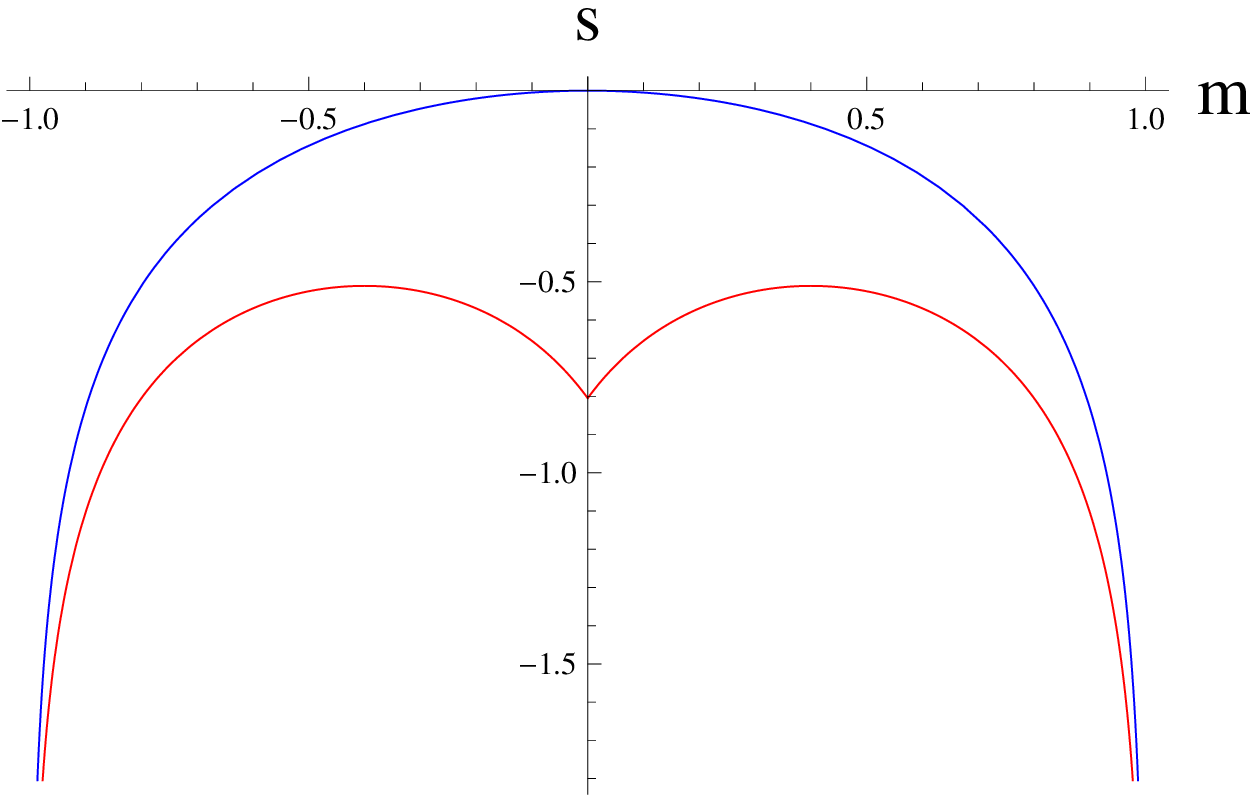}
\includegraphics[width=0.235\textwidth]{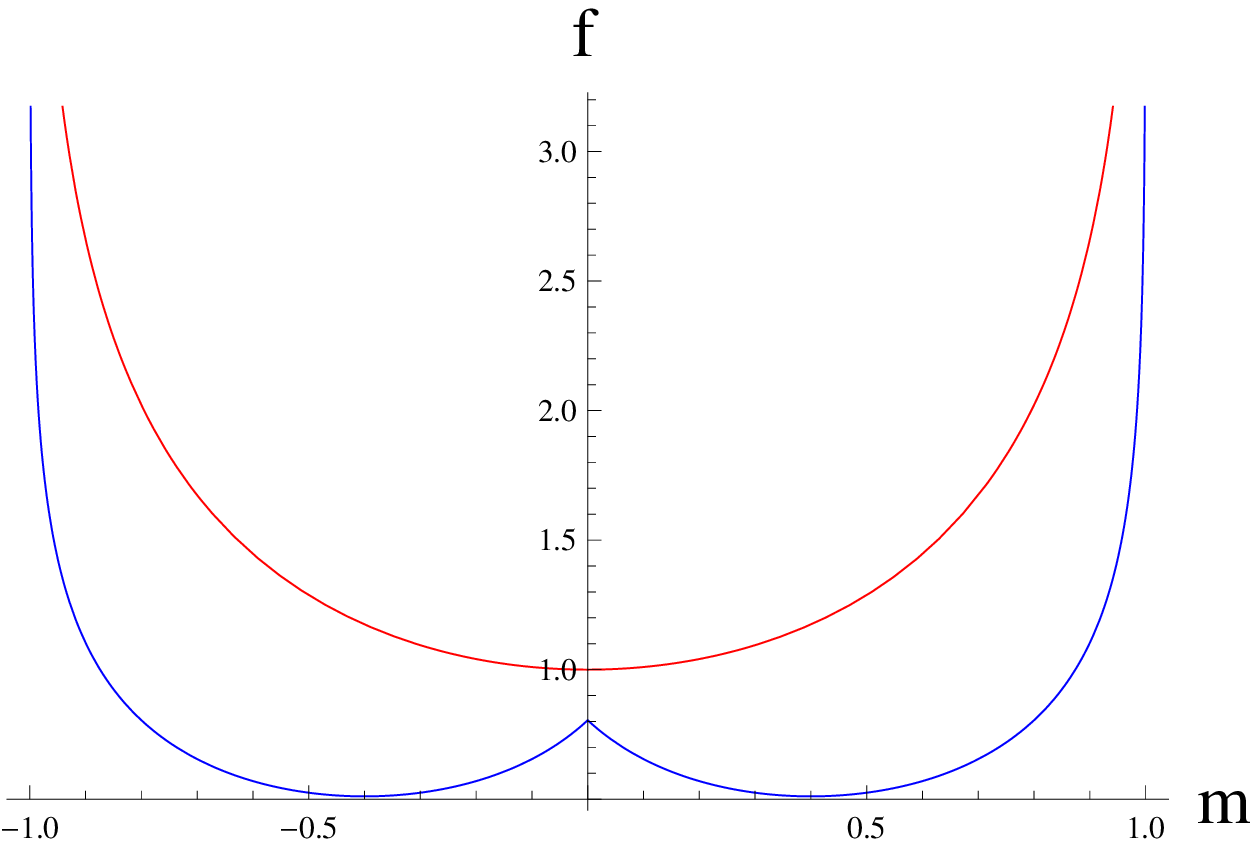}
\caption{Model of Sec. \ref{bs} with $v_c=1$ and $m_0=0.4$. Left: entropy $s(v,m)$ as a function of $m$ at $v=0$ (blue), and at $v=-v_c$ (red). Right: as left for the free energy $f=\left\langle v\right\rangle-Ts(m)$ at $T<T_c$ (blue) and at $T>T_c$ (red). $T_c\simeq 1.96$.}
\label{bs_smfm}
\end{center}
\end{figure}

As for the hypercubic model, the potential takes only the values $-v_c,0$. To calculate the density of states as a function of $m$ we start from $v=v_c$, thus 
\begin{equation}
\omega_N(-v_c,m)=\mu\left(\Sigma_{-v_c,N}\cap\pi_{m,N}\right)=\mu\left(A^+\cup A^-\cap\pi_{m,N}\right).
\end{equation}
$A^+\cup A^-\cap\pi_{m,N}$ is an $(N-1)$-ball whose radius is given by 
\begin{equation}
r(m)=\sqrt{N}\left((1-m_0)^2-(|m|-m_0)^2\right)^{\frac{1}{2}}.
\end{equation}
Thus, the density of states at fixed $m$ is the volume of the $(N-1)$-ball of radius $r(m)$
\begin{equation}
\omega_N(-v_c,m)=\frac{\pi^{\frac{N}{2}}N^{\frac{N-1}{2}}}{\Gamma\left(\frac{N}{2}+1\right)}\left((1-m_0)^2-(|m|-m_0)^2\right)^{\frac{N-1}{2}}.
\end{equation}
Finally, the entropy in the thermodynamic limit is given by 
\begin{eqnarray}
s(-v_c,m)=\lim_{N\to\infty}\frac{1}{N}\ln \omega_N(-v_c,m)\nonumber
\\
=\frac{1}{2}\ln\left((1-m_0)^2-(|m|-m_0)^2\right).
\end{eqnarray}
$s(-v_c,m)$ is plotted in Fig. \ref{bs_smfm}, it shows a relative minimum at $m=0$ and two absolute maxima at $m=\pm m_0$ corresponding to the spontaneous magnetization. The 'strangledness' of $A^+\cup A^-$ has reflected in a non-concave entropy entailing the spontaneous breaking of the $\mathbb{Z}_2$ symmetry.

Now, consider the case $v=0$. 
\begin{eqnarray}
\omega_N(0,m)=\mu\left(\Sigma_{0,N}\cap\pi_{m,N}\right)\nonumber
\\
=\mu\left(\left(B\backslash\left(A^+\cup A^-\right)\right)\cap\pi_{m,N}\right)\nonumber
\\
=\mu\left(\left(B\cap\pi_{m,N}\right)\backslash\left(A^+\cup A^-\cap\pi_{m,N}\right)\right)
\end{eqnarray}
Since the radius of $A^+\cup A^-\cap\pi_{m,N}$ is less than that of $B\cap\pi_{m,N}$, the contribution of the former is vanishing for $N\rightarrow\infty$ with respect to the latter, so that
\begin{eqnarray}
\omega_N(0,m)=\mu\left(B\cap\pi_{m,N}\right)\nonumber
\\
=\frac{\pi^{\frac{N}{2}}N^{\frac{N-1}{2}}}{\Gamma\left(\frac{N}{2}+1\right)}\left(1-|m|\right)^{N-1},
\end{eqnarray}
and finally, the entropy is
\begin{eqnarray}
s(0,m)=\lim_{N\to\infty}\frac{1}{N}\ln \omega_N(0,m)=\ln\left(1-|m|\right).
\end{eqnarray}

The importance of this model is in showing in an discrete and elementary way the generating-mechanism of a $\mathbb{Z}_2$-SBPT. In the general case this occurs in a continuous way giving rise in some cases to a continuous SBPT, i.g. the $\phi^4$ model that we will analyze in next Section.

\section{A possible application: mean-field $\phi^4$ model}
\label{phi4}

We recall the potential of the mean-field $\phi^4$ model with a $\mathbb{Z}_2$ symmetry
\begin{equation}
V=\sum^{N}_{i=1}\left(-\frac{\phi_i^2}{2}+\frac{\phi_i^4}{4}\right)-\frac{J}{2N}\left(\sum^{N}_{i=1}\phi_i\right)^2.
\end{equation}
The model is known to undergo a second-order $\mathbb{Z}_2$-SBPT with classical critical exponents.

\begin{figure}
\begin{center}
\includegraphics[width=0.235\textwidth]{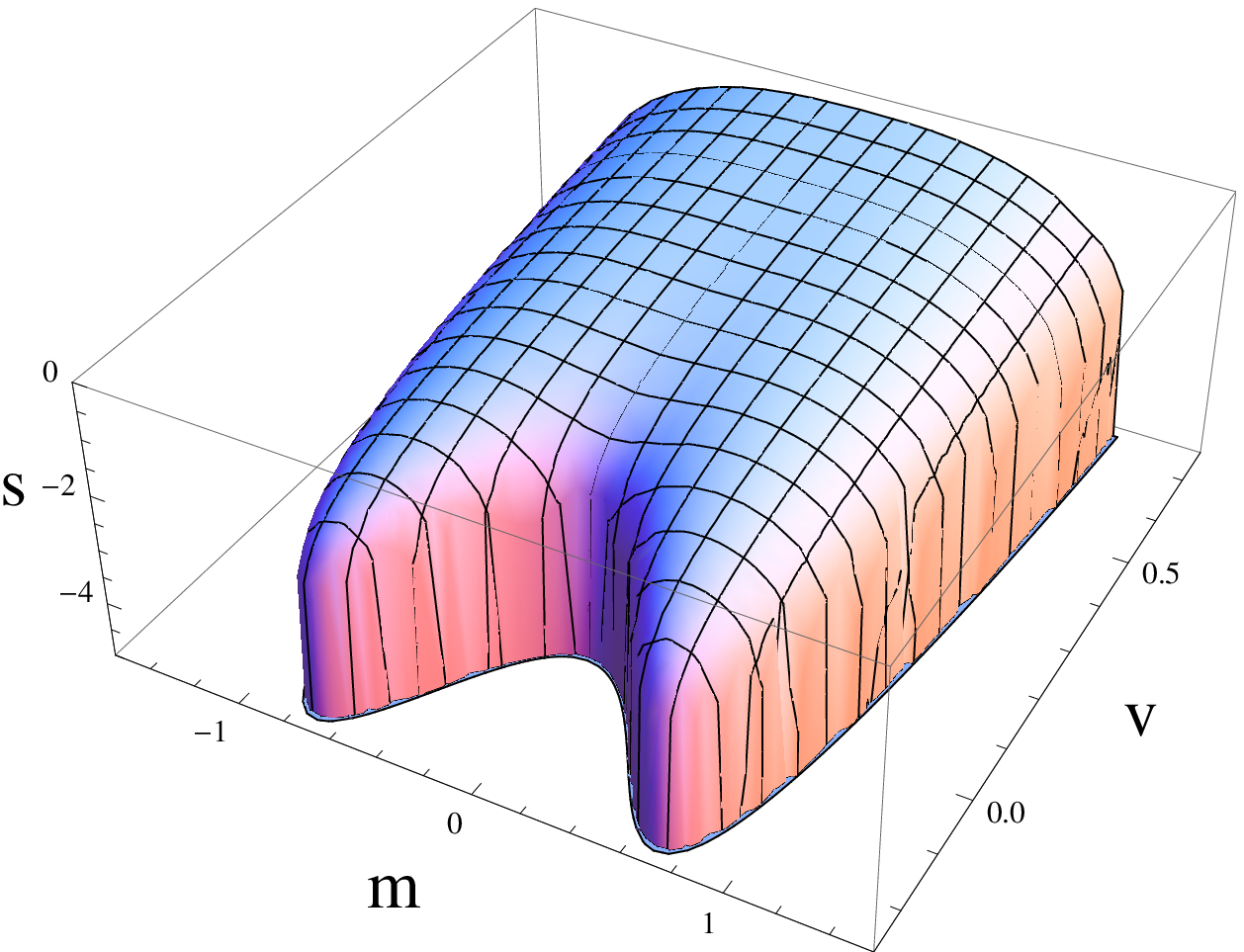}
\includegraphics[width=0.235\textwidth]{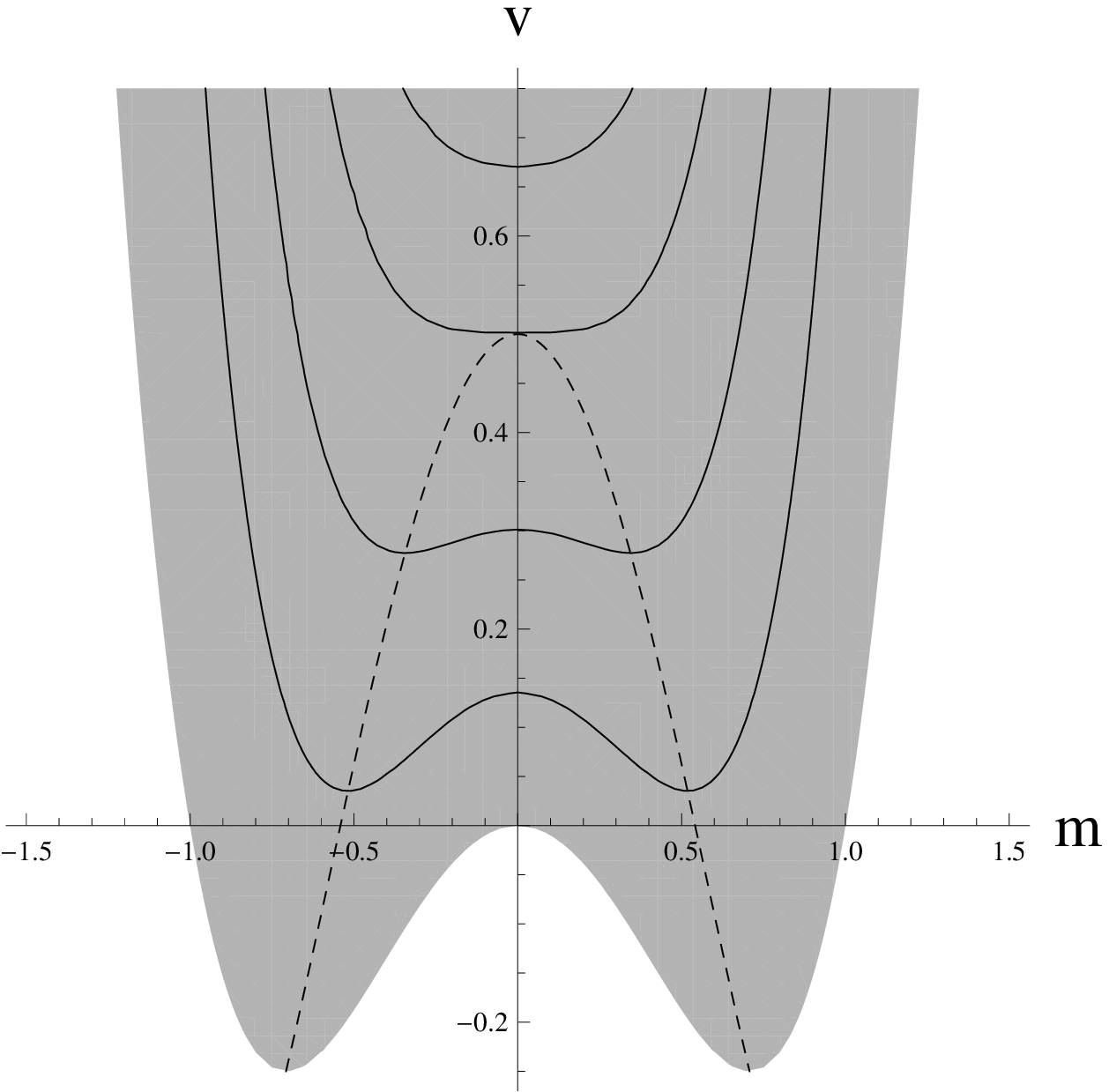}
\caption{Mean-field $\phi^4$ model (\ref{Vphi4}) with coupling $J>0$ (the plots reported here are reproductions only in qualitative agreement with the originals Fig. $5$ in \cite{k} and Fig. $2$ in \cite{hk}, respectively from left to right, to which we refer for a quantitative study). Left: graph of the microcanonical entropy $s(v,m)$ as a function of the potential $v$ and the magnetization $m$. Right: contour plot of $s(v,m)$. The gray hatched area is the domain of $s$, the continuous curves are level sets for some values of $s$, and the dashed curve is the spontaneous magnetization.}
\label{kastner}
\end{center}
\end{figure}

In \cite{hk} the authors have been able to calculate the thermodynamic limit of the microcanonical entropy $s(v,m)$ (see Fig. \ref{kastner}) of the mean-field $\phi^4$ model by large deviations theory. The canonical entropy $\hat{s}(v)$ is obtained by a process of maximization of $s(v,m)$ with respect to $m$
\begin{equation}
\hat{s}(v)=\max_{m}s(v,m).
\label{Vphi4}
\end{equation}
The domain of $s(v,m)$ is a non-convex subset of the plane $(v,m)$, and $s(v,m)$ is a non-concave function, coherently with the long-range interaction of the model. The critical average potential $v_c$ of the SBPT is located in such a way to divide the concave sections $s(v,m)$ at fixed $v$ at $v\ge v_c$ from the non-concave ones at $v<v_c$.

In \cite{aarz,b0,gss} the topology of the $\Sigma_{v,N}$'s hes been exhaustively studied by Morse theory \cite{pettini}. The following three cases have delineated:

\smallskip
(i) $v\in [v_{min},v_t)$, where $v_{min}=-\frac{1}{4}(1+J)^2$ is the absolute minimum of the potential. $v_t$ depends on the coupling constant $J$, and $v_t<-\frac{1}{4}$. The $\Sigma_{v,N}$'s are homeomorphic to the union of two disjoint $N$-spheres. The critical potential of the SBPT may be less than $0$, but $v_c>v_t$ holds for every $J$. 

\smallskip
(ii) $v\in [v_t,0]$. There is a huge amount of critical points growing as $e^N$, and as a consequence of topological changes. We can say that the whole interval $[v_t,0]$ plays the role of a critical $v$-level set, because discriminates between the $\Sigma_{v,N}$'s homeomorphic to two disjoint $N$-spheres from the ones homeomorphic to a single $N$-sphere. In a future paper we will see how it is possible to reduce this critical interval to a single critical $v$-level set containing a single critical point. Furthermore, as $J\rightarrow\infty$, $v_t\rightarrow-\frac{1}{4}^-$.

\smallskip
(iii) $v\in (0,+\infty)$. The $\Sigma_{v,N}$'s are homeomorphic to an $N$-sphere. 

\smallskip
Let us try to interpret this scenario in the light of the results of Sec. \ref{strangled}. 

In the case (i) the hypotheses of theorem 1 in \cite{bc} are satisfied, thus the topology of the $\Sigma_{v,N}$'s implies the $\mathbb{Z}_2$-SB. This is in accordance with $v_c>v_t$ for every $J$, because the magnetization cannot vanish below $v_t$. As showed in Sec. \ref{strangled}, since the theorem in \cite{bc} is a particular case of that given in Sec. \ref{strangled}, also the hypotheses of the latter are satisfied. 

In the case (ii) the hypotheses of the theorem in \cite{bc} are not satisfied, so that only the theorem given in Sec. \ref{strangled} can implies the $\mathbb{Z}_2$-SB, because the $\Sigma_{v,N}$'s may be dumbbell-shaped below $v_c$ and not anymore above $v_c$ (if $v_c<0$) independently on their intricate topology.

Finally, the same of the case (ii) holds for the case (iii), with the non-significant difference that the $\Sigma_{v,N}$'s are all diffeomorphic to an $N$-sphere.

$\Sigma_{v_c,N}$ plays the role of critical $v$-level set in the sense that it separates the dumbbell $\Sigma_{v,N}$'s from that which are not dumbbell-shaped. In general, for more precision, we aspect that at fixed $N$ the critical $\Sigma_{v,N}$ in the above-specified sense is not located exactly at $v_c$, but there may exist a sequence of critical $\Sigma_{v_c^N,N}$ such that $v_c^N\rightarrow v_c$ for $N\rightarrow\infty$. Further analytic and numerical studies may check this conjecture. 

\section{Concluding remarks}

In this paper we have highlighted the generating-mechanism for $\mathbb{Z}_2$-SBPTs in Hamiltonian systems. The mechanism is based on a particular shape of the $v$-level sets that we have defined dumbbell-shaped. A dumbbell $v$-level set entails the spontaneous ergodicity breaking, whence the SBPT. Since in the thermodynamic limit the the $\left\langle v\right\rangle(T)$-level sets is the unique level set accessible to the RP of the system, we can calculate the spontaneous magnetization by the ensemble average on the $\left\langle v\right\rangle(T)$-level set provided that the last is ergodic. But we have shown that if the $\left\langle v\right\rangle(T)$-level set is dumbbell-shaped, then it cannot be ergodic because the RP cannot visit all the regions of the $\left\langle v\right\rangle(T)$-level set for a time proportional to the statistical measure. In other words, a dumbbell $v$-level set has a neck in correspondence of the plane at $0$-magnetization where the probability to find the RP is exponentially suppressed with $N$, so that the neck acts as a topological barrier. We note that, despite we have applied this idea to the canonical ensemble, this mechanism is suitable for application also to the microcanonical ensemble because it is independent on $N$.  

The concept of the dumbbell $v$-level sets can be directly extended also to discrete variables-systems, e.g. the Ising model. The $v$-level sets become discrete sets, so that the statistical measure is replaced by the count of the microstates. 

This question weather a narrow neck can break the symmetry of a system has been pointed out also in the recent paper \cite{gfp}, where the authors have introduced the new concept of asymptotic diffeomorphicity among manifolds: two manifolds can be diffeomorphic at any finite $N$, but not asymptotically diffeomorphic in the limit $N\rightarrow\infty$. In the meanwhile we wait for a rigorous definition of asymptotic diffeomorphicity, we wonder weather this picture may be equivalent to the one of the dumbbell $v$-level sets introduced here, at least for systems with a $\mathbb{Z}_2$ symmetry where dumbbell $v$-level sets are defined. 

Another open question is about short-range systems. In a short-range system the entropy $s(v,m)$ has to be a concave function of $v$, at most non-strictly concave in presence of phase transitions \cite{g,l}. The definition of the dumbbell $v$-level sets we have given is suitable only for long-range systems because it implies a non-concave shape of $s(v,m)$. Anyway, that definition is suitable to be generalized to the short-range case if we assume that the 'strangledness' of the $v$-level sets are maintained for any $N$ but not in the limit $N\rightarrow\infty$. This entails that the entropy must be a non-strictly concave function of $v$, as requested for a short-range system. This means that there is no qualitative difference between short-range and long-range as long as $N$ is far from the thermodynamic limit. In \cite{gfp} the average trapping time of the RP to be located around one of the two average values of the magnetization in the broken phase of the $2D$ $\phi^4$ model with nearest neighbors interactions has been found to increase as a polynomial in $N$. But in a long-range system the trapping time increases exponentially in $N$, so that the difference becomes crucial as $N$ is large enough. This difference may be responsible of the fact that a non-concave entropy in the broken phase of a long-range system is replaced by a non-strictly concave entropy of a short-range system. 

Lastly, we wonder how the picture introduced in this paper for $\mathbb{Z}_2$-symmetric systems may be extended to other symmetry groups, for example $O(n)$ with $n>1$.

\begin{acknowledgments}
I would like to thank Michael Kastner for useful discussion and suggestion.
 \end{acknowledgments}


\begin{thebibliography}{99}

\bibitem{acprz} A. Andronico, L. Casetti, M. Pettini, G. Ruocco, and F. Zamponi \textit{Europhys. Lett.} E \textbf{62} 775 (2003)
A. Andronico, L. Casetti, M. Pettini, G. Ruocco, and F. Zamponi \textit{Phys. Rev.} E \textbf{71} 036152 (2005)
\bibitem{aarz} A. Andronico, L. Angelani, G. Ruocco, and F. Zamponi \textit{Phys. Rev.} E \textbf{70} 041101 (2004)
\bibitem{b1} F. Baroni \textit{arXiv:}1106.3870v4 [cond-mat.stat-mech] (2011)
\bibitem{b0} F. Baroni \textit{arXiv:}1611.03960 (2016)
\bibitem{b} F. Baroni \textit{J. Stat. Mech.} \textbf{P} 08010 (2011) 
\bibitem{bc} F. Baroni, and L. Casetti \textit{J. Phys. A: Math. Gen.} \textbf{39} 529-545 (2006)
\bibitem{bk} T.H. Berlin, and M. Kac \textit{Phys. Rev.} \textbf{86} 821-835 (1952)
\bibitem{cccp} L. Caiani, L. Casetti, C. Clementi, and M. Pettini \emph{Phys. Rev. Lett.} \textbf{79} 4361 (1997)
\bibitem{ccp} L. Casetti, E.G.D. Cohen, and M. Pettini \textit{Phys. Rev. Lett.} \textbf{82} 4160 (1999)
\bibitem{ccp1} L. Casetti, E.G.D. Cohen, and M. Pettini \textit{Phys. Rep.} \textbf{337} 237 (2000)
\bibitem{ccp2} L. Casetti, E.G.D. Cohen, and M. Pettini \textit{Phys. Rev.} E \textbf{65} 036112 (2002)
\bibitem{cfsp} L. Casetti, R. Franzosi, M. Pettini, and L. Spinelli \textit{Phys. Rev.} E \textbf{60} R5009 (1999)
\bibitem{ck} L. Casetti, and M. Kastner \textit{Phys. Rev. Lett.} \textbf{97} 100602 (2006)
\bibitem{ck1} L. Casetti, and M. Kastner \textit{Physica A} \textbf{384} 318 (2007)
\bibitem{ckn} L. Casetti, M. Kastner, and R. Nerattini \textit{J. Stat. Mech.} P07036 (2009)
\bibitem{ccp3} M. Cerruti-Sola, C. Clementi, and M. Pettini \textit{Phys. Rev.} \textbf{61} 5171 (2000)
\bibitem{c} E.G.D. Cohen \textit{Ann. J. Phys.} \textbf{58} (1990)
\bibitem{f} M.E. Fisher, The nature of critical points, in: W. E. Brittin (ed.), \emph{Lectures in Theoretical Physics} (University of Colorado Press, Boulder 1965), Vol. VII, Part c
\bibitem{fps} R. Franzosi, M. Pettini, and L. Spinelli \textit{Phys. Rev. Lett.} \textbf{84} 2774 (2000)
\bibitem{fp} R. Franzosi, and M. Pettini \textit{Phys. Rev. Lett.} \textbf{92} 060601 (2004)
\bibitem{fp1} R. Franzosi, and M. Pettini \textit{Nucl. Phys. B} \textbf{782} 219 (2007)
\bibitem{g} G. Gallavotti \emph{Statistical Mechanics: A Short Treatise} (Springer, New York) (1999)
\bibitem{goldenfeld} N. Goldenfeld \emph{Lectures on Phase Transitions and the Renormalization Group} (Cambridge: Perseusn Publishing, 1992)
\bibitem{gss}  D.A. Garanin, R. Shilling, and A. Scala \textit{Phys. Rev.} E \textbf{70} 036125 (2004)
\bibitem{gfp} M. Gori, R. Franzosi, and M. Pettini, \textit{arXiv:}1602.01240 [cond-mat.stat-mech] (2016)
\bibitem{gm} P. Grinza, and A. Mossa \textit{Phys. Rev. Lett.} \textbf{92} 158102 (2004)
\bibitem{hk} I. Hahn, and M. Kastner \textit{Phys. Rev. E} \textbf{72} 056134 (2005)
\bibitem{hk1} I. Hahn, and M. Kastner \textit{Eur. Phys. J. B} \textbf{50}, 311-314 (2006)
\bibitem{huang} K. Huang \textit{Statistical Mechanics} (John Wiley and Sons, 1987)
\bibitem{ising} E. Ising \textit{Z. Phys.} \textbf{31} 253-258 (1925)
\bibitem{k} M. Kastner \textit{Rev. Mod. Phys.} \textbf{80} 167 (2008)
\bibitem{k1} M. Kastner \textit{Phys. Rev. Lett.} \textbf{93} 150601 (2004)
\bibitem{km} M. Kastner, and D. Metha \textit{Phys. Rev. Lett.} \textbf{100} 160601 (2008)
\bibitem{ks} M. Kastner, and O. Schnetz \textit{Phys. Rev. Lett.} \textbf{107} 160602 (2011)
\bibitem{kss1} M. Kastner, S. Schreiber, and O. Schnetz \textit{Phys. Rev. Lett.} \textbf{99} 050601 (2007)
\bibitem{kss} M. Kastner, O. Schnetz, and S. Schreiber \textit{J. Stat. Mech.} P04025 (2008)
\bibitem{l} O.E. Lanford "Entropy and equilibrium states in classical statistical mechanics," in \emph{Statistical Mechanics and Mathematical Problems}, edited by A. Lenard, Lectures Notes in Physics Vol. 20  (Springer, New York), 1 (1973)
\bibitem{lebowitz} J.L. Lebowitz \textit{Rev. Mod. Phys.} \textbf{71} S346 (1999)
\bibitem{mhk} D. Mehta, J. D. Hauenstein, and M. Kastner \textit{Phys. Rev. E} \textbf{85}, 061103 (2012)
\bibitem{mk} D. Mehta, and M. Kastner \emph{arXiv:}1010.5335v1 [cond-mat.stat-mech]
\bibitem{o} L. Onsager \textit{Phys. Rev.} \textbf{65} 117 (1944)
\bibitem{pettini} M. Pettini \textit{Geometry and Topology in Hamiltonian Dynamics and Statistical Mechanics} (Springer-Verlag New York Inc., 2007)
\bibitem{palmer} R.G. Palmer \textit{Adv. Phys.} \textbf{31} 669 (1982)
\bibitem{p} R.G. Palmer \textit{Adv. Phys.} \textbf{87} 404 (1952)
\bibitem{rs}  A.C. Ribeiro-Teixeira, and D.A. Stairolo \textit{Phys. Rev.} \textbf{70} 016113 (2004)
\bibitem{yl} C.N. Yang, and T.D. Lee \textit{Phys. Rev.} \textbf{87} 404-410 (1952)

\end{thebibliography}
\end{document}